\documentclass[twocolumn,showpacs,preprintnumbers,amsmath,amssymb,showkeys]{revtex4}

\usepackage{epsfig}
\usepackage{dsfont}

\usepackage{graphicx}
\usepackage{dcolumn}
\usepackage{bm}

\newcommand{\beq}{\begin{equation}}
\newcommand{\eeq}{\end{equation}}
\newcommand{\bea}{\begin{eqnarray}}
\newcommand{\eea}{\end{eqnarray}}
\newcommand{\nn}{\nonumber \\}

\newcommand\eqn[1]{(\ref{#1})}      
\newcommand\Eqn[1]{Eq.~(\ref{#1})}  
\newcommand\Fig[1]{Fig.~\ref{#1}}  

\newcommand{\tr}{\hbox{tr}}

\begin{document}


\title{Quantum properties of a non-Gaussian state\\ in the large-$N$ approximation}

\author{F. Gautier}
 \email{fgautier@apc.univ-paris7.fr}
\author{J. Serreau}%
 \email{serreau@apc.univ-paris7.fr}
\affiliation{%
 Astro-Particle and Cosmology\footnote{APC is unit\'e mixte de recherche UMR7164 (CNRS, Universit\'e Paris 7, CEA, Observatoire de Paris).} (APC), University Paris 7 - Denis Diderot\\ 10, rue Alice Domon et L\'eonie Duquet, 75205 Paris Cedex 13, France.
}%

\date{\today}

\begin{abstract}
We study the properties of a non-Gaussian density matrix for an $O(N)$ scalar field in the context of the incomplete description picture. This is of relevance for studies of decoherence and entropy production in quantum field theory. In particular, we study how the inclusion of the simplest non-Gaussian correlator in the set of measured observables modifies the effective (Gaussian) description one can infer from the knowledge of the two-point functions only. We compute exactly the matrix elements of the non-Gaussian density matrix at leading order in a $1/N$-expansion. This allows us to study the quantum properties (purity, entropy, coherence) of the corresponding state for arbitrarily strong nongaussianity.  We find that if the Gaussian and the non-Gaussian observers essentially agree concerning quantum purity or correlation entropy, their conclusion can significantly differ for other, more detailed aspects such as the degree of quantum coherence of the system.
 \end{abstract}

\maketitle


\section{Introduction}
\label{sec:intro}

To completely specify the state of a (quantum) system requires one to perform as many measurements as there are independent degrees of freedom (d.o.f.). This may be difficult in practice for systems with a large number of d.o.f. or even impossible when the latter is infinite, e.g. for (quantum) fields. In fact, one often has to infer the effective state of the system from a restricted amount of information, which may or may not give a good description of the actual state of the system. 

This incomplete description picture \cite{Balian} provides a basis for various studies of entropy production in quantum field theory (QFT) \cite{Calzetta:2003dk,Campo:2008ij}. In this context, one defines an entropy which measures the amount of missing information in the set of measured observables. More recently, similar ideas have been advocated to discuss quantum (de)coherence in QFT \cite{Campo:2008ju, Giraud:2009tn,Koksma:2009wa}. Both issues are of great interest in various contexts such as inflationnary cosmology \cite{Polarski:1995jg, Campo:2008ju}, baryogenesis \cite{Herranen:2008di}, neutrino \cite{Giunti:2002xg} and quark-gluon plasma physics \cite{Muller:2005yu}, or the physics of ultracold quantum gases \cite{Graham}.

First principle calculations of the real-time dynamics of entropy production and quantum decoherence in QFT have been performed recently in Refs. \cite{Giraud:2009tn,Koksma:2009wa} in the case where the set of measured observables is restricted to Gaussian, two-point, correlators of the fields. As a result of the nonlinear evolution, information flows from the subset of Gaussian correlations to -- unmeasured -- higher correlations. In \cite{Giraud:2009tn}, it was demonstrated in the context of a self-interacting $O(N)$ scalar field that, even in the absence of an environment, a pure, coherent quantum state eventually appears as mixed and decohered, and at late times even as a thermal state, to the observer restricted to Gaussian correlators -- we shall call the later the Gaussian observer -- even though the time evolution is unitary. The case of a field coupled to an environment has been considered in \cite{Koksma:2009wa}, with similar conclusions.

Very interesting questions in this context concern the inclusion of higher correlation functions in the set of measured observables. How much information one recovers by measuring the first non-Gaussian correlators? Is the spread of information in the space of correlation functions uniform? Or is it mainly concentrated in a subset of relevant observables? Besides the above-mentionned applications, such questions are of general interest for the understanding of nonequilibrium dynamics and thermalization in QFT \cite{Berges:2004vw}.

In \cite{Koksma:2010zi} Koksma, Prokopec and Schmidt have studied in detail the question of the correlation entropy for various deformations of Gaussian density matrices in scalar field theories. Here, we consider the non-Gaussian density matrix one obtains in the incomplete description picture by including the simplest non-Gaussian correlator -- namely the field four-point function -- in the set of measured observables. We study how the inclusion of this bit of information modifies the conclusions one would infer from the knowledge of  the two-point functions only. 

We find that, depending on what they ask, the Gaussian and non-Gaussian observers may arrive at very different conclusions. 
In particular, if they essentially agree concerning global properties such as the quantum purity or the entropy of the system, whatever the strength of the nongaussianity, they can be led to very different conclusions concerning other, more detailed aspects of the quantum state, such as the degree of quantum coherence or the shape of the probability distributions of various observables. 
For instance, we exhibit situations where the Gaussian observer concludes to an essentially thermal state, whereas the non-Gaussian observer sees a highly coherent quantum state.

Specifically, we consider a situation with space translation invariance, where the fields can be described by their Fourier modes and we assume our observers measure Gaussian and non-Gaussian correlators of separate modes. In this case, the effective density operator is a direct product of individual density operators for each Fourier mode. We focus here on one of these individual operators -- that of the zero mode to be precise. This is essentially equivalent to considering the density operator of a quantum mechanical model with one degree of freedom. 

In principle this could be computed exactly with numerical techniques. Instead, we want to have as much analytical understanding as possible as well as simple practical tools which can be easily implemented in future nonequilibrium calculations \cite{workinprogress}. For this purpose, we consider an $N$-uplet of scalar fields in an $O(N)$-symmetric state and compute the desired properties in the limit $N\to\infty$, for which we get semi-analytical results. Moreover, such a nonperturbative approach allows us to study the case of arbitrarily large nongaussianities.

We describe the reduced density matrix inferred by the ``non-Gaussian'' observer, in Section \ref{sec:dm}. Some global, basis-independent, properties of the corresponding quantum state, such as correlation entropy or quantum purity, are analyzed in Section \ref{sec:entropy}. In Section \ref{sec:matrixel}, we present the calculation (new to our knowledge) of the matrix elements of the non-Gaussian density operator in the limit $N\to\infty$. We analyze the detailed properties of the quantum state, compute the Wigner function on phase space, and study the issue of quantum coherence in Section \ref{sec:Wigner}. A number of appendices are devoted to technical details.

\section{The reduced density matrix}
\label{sec:dm}

We consider $N$ pairs of canonically conjugate scalar field operators $(\hat\varphi_a,\hat\pi_a), a=1,\ldots,N$. We assume a $O(N)$-symmetric state with given Gaussian correlators 
\beq
 \langle\hat\pi_a\hat\pi_b\rangle=\delta_{ab}K,\,\langle\hat\varphi_a\hat\varphi_b\rangle=\delta_{ab}F,\, \langle\hat\varphi_a\hat\pi_b+\hat\pi_b\hat\varphi_a\rangle=2\delta_{ab}R,
\eeq 
and non-Gaussian four-field correlator
\beq
 \langle\hat\varphi_a\hat\varphi_b\hat\varphi_c\hat\varphi_d\rangle=(\delta_{ab}\delta_{cd}+\delta_{ac}\delta_{bd}+\delta_{ad}\delta_{bc})\left[F^2+\frac{C_4}{N}\right].
\eeq
Here $C_4/N$ denotes the non-trivial, connected contribution. The least biased effective density matrix $D$ one can infer from the knowledge of these correlators is the one which is consistent with the measured observables, i.e. $\langle \varphi^2\rangle=\tr[D\,\varphi^2]$ etc., and which maximizes the amount of missing information, measured by the von Neumann entropy $S=-\tr [D\ln D]$. It is given by \cite{Balian}
\beq
\label{eq:dmatrix}
 D = \frac{1}{Z}\exp(-{\cal F})\quad{\rm with}  \quad Z=\tr\exp(-{\cal F})\,,
\eeq
where (we extract an explicit factor $N$ for the purpose of the $1/N$-expansion)
\beq
\label{eq:F}
 {\cal F}=A\hat\pi_a\hat\pi_a+B\hat\varphi_a\hat\varphi_a+C(\hat\varphi_a\hat\pi_a+\hat\pi_a\hat\varphi_a)+\frac{\eta}{N}(\hat\varphi_a\hat\varphi_a)^2\!.\,
\eeq
The parameters  $A$, $B$, $C$ and $\eta$ must be adjusted so that the correlators $K$, $F$, $R$ and $C_4$ obtained from $D$ agree with their measured values. For instance, they can be obtained from the following relations
\beq
\label{eq:coeff}
 K=-\frac{\partial_A\ln Z}{N}\,,\,\,
 F=-\frac{\partial_B\ln Z}{N}\,,\,\,
 R=-\frac{\partial_C\ln Z}{2N}
\eeq
and
\beq
\label{eq:deta}
 F^2+\frac{C_4}{N}=-\frac{\partial_\eta\ln Z}{N+2}\,.
\eeq

\subsection{Large-$N$ limit}

Following standard procedures, the partition function $Z$ can be given the path integral representation
\beq
\label{eq:funcint}
 Z=\int_{\rm periodic}\!\!{\cal D}\varphi{\cal D}\pi\,e^{\int_0^1 d\tau\,\left\{i\dot\varphi_a(\tau)\pi_a(\tau)-{\cal F}(\varphi(\tau),\pi(\tau))\right\}}
\eeq
where $\dot\varphi\equiv d\varphi/d\tau$ and where the functional integral is to be performed on periodic configurations $\varphi_a(0)=\varphi_a(1)$ \footnote{One must pay attention to the ordering procedure. A careful analysis shows that no subtleties arise here.}. Performing the $\pi$ integration, one gets
\beq
 Z={\cal N}(A^N)\int_{\rm per.}\!\!{\cal D}\varphi\,e^{\int_0^1\! d\tau\,\left\{\varphi_a\left(\frac{1}{4A}\frac{d^2}{d\tau^2}-B'\right)\varphi_a-\frac{\eta}{N}(\varphi_a\varphi_a)^2\right\}}
\eeq
where we used the fact that, for periodic field configurations, $\int \!d\tau\,\varphi_a\dot\varphi_a\!=\!0$. Here $B'=B-C^2/A$
and, up to an infinite $A$-independent normalization factor (note that ${\cal N}(AB)={\cal N}(A){\cal N}(B)$)
\beq
 {\cal N}(A)=\int{\cal D}\chi\,e^{-\int_0^1d\tau\,A\chi^2} \propto e^{-\frac{1}{2}{\rm Tr}\ln A}
\eeq
where ${\rm Tr}$ denotes a functional trace.
 
Making use of the standard trick \cite{ZJ}
\beq
\label{eq:trick}
 \int{\cal D} \chi\,e^{\int_0^1 \!d\tau\,\left\{-\frac{N}{4\eta}\chi^2+i\chi\,\varphi_a\varphi_a\right\}}=
 {\cal N}\left(\frac{N}{4\eta}\right)e^{-\frac{\eta}{N}\!\int_0^1\!d\tau\,(\varphi_a\varphi_a)^2}
\eeq
we can perform the $\varphi$-integration to get, up to an irrelevant factor 
\beq
 Z\propto{\cal N}\left(\frac{\eta}{N}\right)\int{\cal D}\chi\, e^{-\frac{N}{2}{\rm Tr}\,{\rm Ln}G^{-1}(\chi)-\frac{N}{4\eta}\int_0^1d\tau\,\chi^2},
\eeq
where
\beq
\label{eq:green}
 G^{-1}(\chi;\tau,\tau')=\left(-\frac{d^2}{d\tau^2}+4A[B'-i\chi(\tau)]\right)\delta(\tau-\tau').
\eeq
This is the starting point for the standard $1/N$-expansion \cite{ZJ}. The limit $N\to\infty$ is given by the saddle point approximation: One has, up to an irrelevant additive constant,
\beq
\label{eq:partition}
 \frac{\ln Z}{N}=-\frac{1}{2}{\rm Tr}\,{\rm Ln}G^{-1}(\bar\chi)-\frac{\bar\chi^2}{4\eta}\,,
\eeq
where $\bar\chi$ is given by the saddle-point equation 
\beq
\label{eq:gap}
 \bar\chi=-\left.\eta\frac{\delta{\rm Tr}\,{\rm Ln}G^{-1}(\chi)}{\delta\chi(\tau)}\right|_{\chi(\tau)=\bar\chi}=4iA\eta {\rm Tr}\,G(\bar\chi)\,.
\eeq
For constant $\bar\chi$, the functional trace on RHS can be evaluated by standard techniques (see the details in Appendix \ref{appFT1}) and \Eqn{eq:gap} can be rewritten as
\beq
\label{eq:gap2}
 \frac{z^2-z_0^2}{\xi}=\frac{1}{z\tanh(z/2)}\,,
\eeq
where $\xi=8A^2\eta$, $z_0=2\sqrt{AB'}$, $z=2\sqrt{A(B'-i\bar\chi)}$. We mention that $z_0^2$ plays the role of a mass squared-like  parameter in the Gaussian case, governing the width of the $\varphi$-distribution and is constrained to be positive. In the non-Gaussian $z^2$ plays the role of a renormalized mass parameter. In Appendix \ref{appS}, we show that only real solutions, i.e.  $z^2>0$, are physically allowed. One easily sees that \Eqn{eq:gap2} always admits one and only one real solution, which can be obtained numerically. Note that one recovers the Gaussian result $z^2=z_0^2$ in the limit $\eta\to 0$. Notice also that, for $\eta\neq0$, $z_0^2$ can be either positive or negative. 

Expressions of the correlators $K$, $F$, $R$ in the limit $N\to\infty$ can be directly obtained from \eqn{eq:partition}, using \eqn{eq:coeff}. The connected correlator $C_4/N$ is ${\cal O}(1/N)$ and necessitates a more involved calculation. The latter is detailed in Appendix \ref{appFT}. In the limit $N\to\infty$, one obtains the parameters of \eqn{eq:F} in terms of the correlators as \footnote{Note that the Heisenberg inequality (see e.g. \cite{Campo:2008ju,Giraud:2009tn}) $FK-R^2\ge1/4\Leftrightarrow\tanh^2(z/2)>0$ indeed implies that $z^2>0$ as also proven in Appendix \ref{appS}.}
\beq
\label{eq:parameters}
 A=\kappa F\,,\quad B = \kappa K - 2\eta F\,,\quad C = -\kappa R\,,
\eeq
where 
\beq
 \kappa=\frac{\ln(1+1/n)}{2n+1}\quad{\rm and}\quad n+{1\over 2}=\sqrt{FK-R^2}\,.
\eeq
It is remarkable that only the parameter $B$ gets modified by nongaussiantiy. This is a consequence of both our choice of non-Gaussian correlator and the large-$N$ approximation.
Note that $n=(e^{z}-1)^{-1}$, with $z$ the solution of the gap equation above.
The value of $\eta$ can be obtained from the relation (see Appendix \ref{appFT1})
\bea
\label{eq:c4}
 &&\hspace{-.8cm}\frac{C_4}{2F^2} = -\frac{x}{2n+1}\left\{\,\frac{2}{x}\left[f(1)-f\left(\!\sqrt{1+x}\,\right)\right]\right.\nn
 &&\qquad\quad+\left.\frac{1}{\ln(1\!+\!1/n)}\left[\frac{\zeta^2}{1+\zeta x}-\frac{1}{1+x}\right]\right\}\!.
\eea
where we defined 
\beq
 \zeta=1+2\kappa n(n+1)\quad{\rm and}\quad x=\frac{\eta}{\kappa}\left(\frac{F}{n+1/2}\right)^2
\eeq
and (notice that $f(1/2)=2n+1$)
\beq
 f(y)=\frac{1}{2y\tanh[y\ln(1\!+\!1/n)]}=\frac{1}{2y}\frac{(1\!+\!n)^{2y}+n^{2y}}{(1\!+\!n)^{2y}-n^{2y}}\,.
\eeq
In the perturbative -- near Gaussian -- regime, small values of the connected correlator $C_4$ as compared to the disconnected contribution $F^2$ corresponds to small values of $x$. In this regime, one gets the linear relation
\beq
 \frac{C_4}{2F^2} =-\frac{x}{2(2n+1)^2}\Big(1+2n(n+1)\left(3\zeta+2\right)\Big)\,.
\eeq
In contrast, for arbitrarily large values of $x$, the ratio $C_4/F^2$ saturates due to non-perturbative effects. We see that $-1\le C_4/2F^2\le0$. It is interesting to notice that the extremal value $C_4=-2F^2$ actually exactly cancels the corresponding (in the $1/N$-expansion) connected contribution to the correlator $\langle(\varphi_a\varphi_a)^2/N\rangle=NF^2+(2F^2+C_4)+{\cal O}(1/N)$.
To separate between the perturbative (near Gaussian) and non-perturbative (strongly non-Gausssian) regimes, we note that the Gaussian square-mass parameter $z_0^2=\ln^2(1\!+\!1/n)(1-2x)$ turns negative for $x> 1/2$. We expect a qualitative change in the properties of the density operator when this happens.

\begin{figure}[t!]  
 \epsfig{file=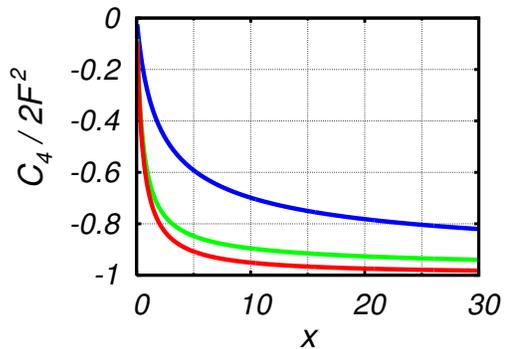,width=7.cm}
 \caption{\label{fig:c4} 
 The ratio $C_4/2F^2$ as function of $x$ for $n=0$ (blue), $n=1$ (green) and $n=10$ (red). The red and blue curves are well described by the limiting functions \eqn{eq:ninf} and \eqn{eq:n0} respectively.}
\end{figure}

We note that, as a function of $x$ the ratio $C_4/2F^2$ is bounded from below by the large-$n$ limit:
\beq
\label{eq:ninf}
 \left.\frac{C_4}{2F^2}\right|_{n\gg1}=-\frac{2x}{1+2x}
\eeq
and from above by the small-$n$ limit:
\beq
\label{eq:n0}
 \left.\frac{C_4}{2F^2}\right|_{n\ll1} = -\frac{x}{1+x+\sqrt{1+x}}.
\eeq
We plot the ratio $C_4/2F^2$ as a function of $x$ for various values of $n$ in \Fig{fig:c4}.

Finally we mention that, using $\bar\chi=2i\eta F$, one has
\beq
\label{eq:lnz}
 \frac{\ln Z}{N}=\ln \sqrt{n(n+1)}+\frac{x}{4}\kappa\left(2n+1\right)^2
\eeq
The first term is the standard Gaussian contribution and the second one the non-Gaussian correction. We now turn to the analysis of the properties of the density matrix \eqn{eq:dmatrix}.

\section{Correlation entropy, quantum purity}
\label{sec:entropy}

For the Gaussian observer $n$ is the only intrinsic -- basis independent -- parameter which characterizes the state of the system. Any basis independent observable, such as correlation entropy or quantum purity, is fully determined by the value of $n$. In this section, we analyze this kind of global intrinsic observables in the non-Gaussian case.

\subsection{Correlation entropy}

The correlation entropy \cite{Balian,Calzetta:2003dk} measures the amount of missing information, i.e. the amount of information not contained in the subset of measured observables. It is defined as  
\beq
 S=-\tr (D\ln D)=\ln Z+\langle{\cal F}\rangle
\eeq
and is easily evaluated in the large-$N$ approximation, using \eqn{eq:F}, \eqn{eq:parameters} and \eqn{eq:lnz}. One finds that the contribution from non-Gaussian terms exactly cancel and one is left with the Gaussian expression \footnote{This is a known property of the large-$N$ approximation, see e.g. \cite{Blaizot:1999ip}.}: 
\beq
 \frac{S}{N}=(n+1)\ln (n+1)-n\ln n\,.
\eeq
This shows that the inclusion of the correlator $C_4$ does not bring any information, in the sense of that measured by entropy. This may seem surprising at first sight since there is clearly some ``information'' in the nongaussianity. But in fact, entropy is a one number which measures the global amount of (missing) information, . There can clearly be an infinity of different density operators having the same entropy.

\subsection{Quantum purity}

The quantity $P=\tr D^2\le1$ measures the quantum purity of the state described by the density operator $D$. It can be computed as 
\beq
 P=\tr D^2=\frac{\tr\, e^{-2{\cal F}}}{(\tr \,e^{-{\cal F}})^2}=\frac{Z(2A,2B,2C,2\eta)}{Z^2(A,B,C,\eta)}
\eeq
where the numerator is obtained as in the previous section by replacing all parameters by twice their values. Writing the solution of the corresponding saddle-point equation $z(2A,2B,2C,2\eta)=2\tilde z$, the latter can be written
\beq
\label{eq:gap2tilde}
  \frac{\tilde z^{2}-z_0^2}{\xi}=\frac{1}{\tilde z\tanh \tilde z}\,.
\eeq

\begin{figure}[t!]
 \epsfig{file=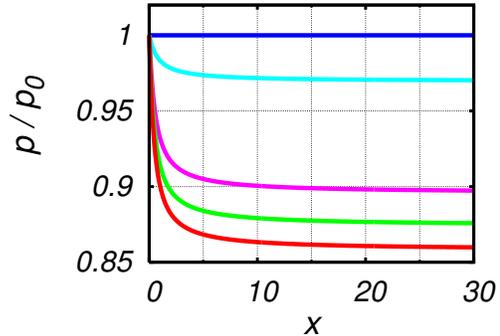,width=7.cm}
 \caption{\label{fig:purity} 
 The ratio of non-Gaussian vs. Gaussian purities $p/p_0$ as a function of the nongaussianity $x$ for, from top to bottom, $n=0$, $0.1$, $0.5$, $1$ and $10$. The latter case is well-described by the $n\gg1$ limit, \Eqn{eq:limit}.}
\end{figure}

In the large-$N$ limit, the $N$ field components are essentially identical and independent degrees of freedom and the total purity can be written as $P= p^N$ where $p$ can be viewed as the individual purity of each d.o.f. \footnote{Of course any small deviation from $p=1$ results in a large deviation from $P=1$ as a mere consequence of the large number of independent d.o.f. The relevant quantity here is the individual purity $p$.}. Introducing $\tilde n=(e^{\tilde z}-1)^{-1}$ and using \Eqn{eq:lnz}, we get, after some calculations,
\beq
 p=\frac{1}{2\tilde n+1}\,\frac{\tilde n(\tilde n+1)}{n(n+1)}\,e^{-{1\over2}x\kappa\left(2n+1\right)^2\left\{1-\frac{\kappa^2}{\tilde \kappa^2}\left[\frac{1+2\tilde n(\tilde n+1)}{1+4\tilde n(\tilde n+1)}\right]^2\right\}}
\eeq
where $\tilde\kappa=\ln(1+1/\tilde n)/(2\tilde n+1)=\tilde z\tanh(\tilde z/2)$. Notice that, for real positive $x$, $\tanh x>\tanh(x/2)$ and, therefore, $\tilde z<z$, which implies $\tilde n>n$ and $\tilde \kappa>\kappa$. Notice also that purity only depends on the parameters $n$ and $x$.

We check that we recover the usual result \cite{Giraud:2009tn} in the Gaussian case ($x=0$), for which $z=\tilde z=z_0$:
\beq
 p_{0}=\frac{1}{2n+1}.
\eeq
The pure Gaussian state corresponds to $n\ll1$. A simple analysis shows that the case  $n\ll1\Leftrightarrow z\gg1$ is equivalent to $\tilde n\ll1\Leftrightarrow \tilde z\gg1$. In this situation, one gets, from Eqs. \eqn{eq:gap2} and \eqn{eq:gap2tilde}, $\tilde z\approx z$ and thus 
\beq
 p\approx p_{0}\approx 1 \quad {\rm for}\quad n\ll1 .
\eeq
We conclude that both the Gaussian and the non-Gaussian observers agree on the purity of the system when the latter is pure. 

The other extreme is $n\gg1$, which implies $\tilde n\gg1$. In this case, one has $z\approx 1/n\ll1$ and $\tilde z\approx 1/\tilde n\ll1$. The saddle point equation \eqn{eq:gap2tilde} simplifies to a quadratic equation for $\tilde z^2$ and one gets, after some simple calculations:
\beq
 {\tilde n^2\over n^2}\approx\frac{2}{1-2x+\sqrt{1+4x^2}}
\eeq
and
\beq
\label{eq:limit}
 {p\over p_{0}}\approx {\tilde n\over n}\,e^{-x\left(1-\frac{\tilde n^4}{4n^4}\right)}\quad{\rm for}\quad n\gg1.
\eeq
One easily checks that $1\ge p/p_0>\sqrt{2/e}\approx 0.86$, the two limits corresponding to the Gaussian ($x=0$) and the extreme non-Gaussian ($x\gg1$) cases respectively. We conclude that both observers essentially agree on the degree of quantum purity of the system whatever the value of $n$ and the strength of the nongaussianity. This is demonstrated on \Fig{fig:purity} which shows the individual purity $p$ as a function of the nongaussianity $x$ for given Gaussian correlators, i.e. for fixed $n$. 

Thus we see that both the correlation entropy and the quantum purity are pretty insensitive to the degree of nongaussiantiy of the system \footnote{In Ref. \cite{Koksma:2010zi}, Koksma et al. consider different deformation of a Gaussian density matrix which result in modifications of the entropy. However, their calculation being restricted to small deformations, these modification are also small. It would be interesting to study how these entropy changes evolve as a function of (large) deformations for fixed Gaussian correlators.}. Such global observables are not sensitive to high moments of the underlying distribution and thus do not contain enough information to qualitatively distinguish between the Gaussian and non-Gaussian cases. 

Still, as explained above, we expect substantial, qualitative changes in the detailed properties of the density operators inferred by our two observers. In the next section, we compute the detailed structure of the density operator in the large-$N$ limit and study how it evolves as a function of the nongaussianity. 

\section{Matrix elements in the limit $N\to\infty$}
\label{sec:matrixel}

We now wish to compute the matrix elements of the density operator in various basis. The ``position'' basis, i.e. the basis of eigenvalues of the field operators, turns out to be particularly suited to take the $N\to\infty$ limit. We start with the general functional integral formula: 
\beq
\label{eq:matel}
 \langle\varphi_2|e^{-{\cal F}}|\varphi_1\rangle=\!\int_{\varphi_1}^{\varphi_2}\!{\cal D}\varphi{\cal D}\pi\,e^{\int_0^1 d\tau\,\left\{i\dot\varphi_a\pi_a-{\cal F}(\varphi,\pi)\right\}}
\eeq
where $|\varphi\rangle\equiv\prod_a|\varphi_a\rangle$. Here the functional integral is to be performed on field configurations such that $\vec\varphi(0)=\vec\varphi_1$ and $\vec\varphi(1)=\vec\varphi_2$, where arrows denote $O(N)$ vectors. Note that, because of $O(N)$ symmetry, the matrix element \eqn{eq:matel} only depend on the invariants $\varphi_1^2$, $\varphi_2^{2}$ and $\vec\varphi_1\cdot\vec\varphi_2$. It is useful to introduce the rescaled quantities \footnote{We consider typical $O(N)$ vectors: $u\sim v\sim w\sim 1$.}
\beq
\label{eq:uvw}
 u^2=\frac{\varphi^2}{NA}\,,\quad v^2=\frac{s^2}{4NA}\quad{\rm and} \quad w=\frac{\vec \varphi\cdot\vec s}{2NA}\,,
\eeq
where
\beq
 \vec \varphi=\frac{\vec \varphi_2+\vec \varphi_1}{2}\quad{\rm and} \quad \vec s=\vec \varphi_2-\vec \varphi_1
\eeq

Using \Eqn{eq:trick}, we write
\beq
  \langle\varphi_2|e^{-{\cal F}}|\varphi_1\rangle={\cal N}\left(\frac{4\eta}{N}\right)\int{\cal D}\chi\,e^{-N\left(S_{\varphi_2,\varphi_1}[\chi]+\int_0^1d\tau\,\frac{\chi^2}{4\eta}\right)},
\eeq
where we defined
\beq
\label{eq:S}
 e^{-NS_{\varphi_2,\varphi_1}[\chi]}=\!\int_{\varphi_1}^{\varphi_2}\!{\cal D}\varphi{\cal D}\pi\,e^{\int_0^1 d\tau\,\left\{i\dot\varphi_a\pi_a-{\cal F}_{\rm q}(A,B-i\chi,C)\right\}},
\eeq
with the quadratic form 
\beq
 {\cal F}_{\rm q}(A,B,C)=A\pi_a\pi_a+B\varphi_a\varphi_a+C(\varphi_a\pi_a+\pi_a\varphi_a).
\eeq
As for the calculation of the partition function above, the large-$N$ limit is given by the saddle-point contribution:
\bea
 &&\hspace{-1.cm}\langle\varphi_2|e^{-{\cal F}}|\varphi_1\rangle=e^{-N\left[S_{\varphi_2,\varphi_1}[\bar\chi(u^2,v^2)]+\frac{\bar\chi^2(u^2,v^2)}{4\eta}\right]}\nn 
 \label{eq:largeN}
 &&=\langle\varphi_2|e^{-{\cal F}_{\rm q}(A,B-i\bar\chi(u^2,v^2),C)}|\varphi_1\rangle e^{-\frac{N}{4\eta}\bar\chi^2(u^2,v^2)},
\eea
where the saddle-point $\bar\chi(u^2,v^2)$, given by 
\beq
\label{eq:saddle}
 \frac{\bar\chi(u^2,v^2)}{2\eta}=-\left.\frac{\delta S_{\varphi_2,\varphi_1}[\chi]}{\delta\chi(\tau)}\right|_{\chi(\tau)\equiv\bar\chi(u^2,v^2)},
\eeq
only depends $u^2$ and $v^2$ defined in \Eqn{eq:uvw}, as shown below.

We note that, using \eqn{eq:S}, the saddle-point equation \eqn{eq:saddle} can be rewritten as
\beq
\label{eq:saddle2}
  \frac{\bar\chi(u^2,v^2)}{2\eta}\!=\!\left.\frac{1}{iN}\frac{\partial}{\partial B}\ln\langle\varphi_2|e^{-{\cal F}_{\rm q}(A,B,C)}|\varphi_1\rangle\right|_{B\to B-i\bar\chi(u^2,v^2)}\!\!\!.
\eeq
Therefore, we find that, in the limit $N\to\infty$, the matrix elements of the non-Gaussian operator $\exp(-{\cal F})$ can be entirely expressed in terms of those of the Gaussian operator $\exp(-{\cal F}_q)$. The latter are well-known:
\beq
\label{eq:latter}
 \langle\varphi_2|e^{-{\cal F}_{\rm q}}|\varphi_1\rangle=\frac{1}{A^{N\over2}}e^{-N\left\{F_0(z_0)+F_u(z_0)u^2+F_v(z_0)v^2+2iCw\right\}},
\eeq
where $z_0$ has ben defined in \Eqn{eq:gap2} and  $u,v,w$ in \Eqn{eq:uvw} and where
\bea
 F_0(z)&=&-\ln\sqrt{\frac{z}{4\pi\sinh(z)}},\\
 \label{eq:fu}
 F_u(z)&=&\frac{z}{2}\tanh\left({z\over2}\right),\\
 \label{eq:fv}
 F_v(z)&=&\frac{z/2}{\tanh(z/2)}.
\eea

We emphasize that Eqs. \eqn{eq:largeN} and \eqn{eq:saddle2} in fact hold in any basis and, more generally for any linear combination of matrix elements \footnote{Of course the dependence of the saddle-point $\bar\chi$ in the states at hand may be more complicated than here, but the general structure of the equations remains the same.}. For instance,  Eqs. \eqn{eq:partition} and \eqn{eq:gap} can be obtained as a particular case when one replace $\langle\varphi_2|e^{-{\cal F}_q}|\varphi_1\rangle\to\tr \,e^{-{\cal F}_q}$ in both Eqs. \eqn{eq:largeN} and \eqn{eq:saddle2}. Similarly, we mention that similar equations can be obtained for the Wigner function \eqn{eq:wigner} to be discussed below.

\begin{figure}[t!]
 \epsfig{file=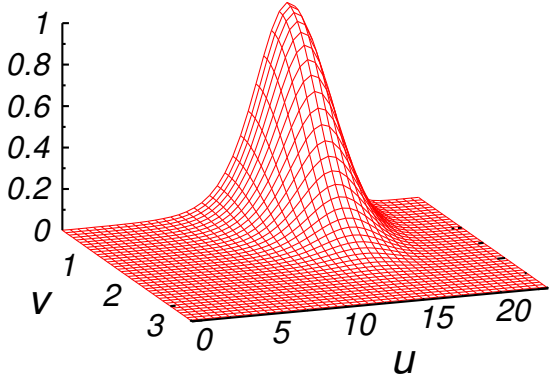,width=10.cm}
 \caption{\label{fig:camel} 
 The function $ d(u^2,v^2)$, normalized to its maximum value in the $(u,v)$-plane for $n=10$ and $x=15$.}
\end{figure}

The position basis employed here presents the advantage that the solutions of \Eqn{eq:saddle2} are particularly simple. Notice first that the phase of $\langle\varphi_2|e^{-{\cal F}_q}|\varphi_1\rangle$ in \eqn{eq:latter} is the only $w$-dependent term and is independent of $B$. Thus this term does not contribute to the saddle-point equation, which implies, first, that $i\bar\chi(u^2,v^2)$ is real and, second, that it only depends on $u^2$ and $v^2$. In terms of the variable $z(u^2,v^2)=2\sqrt{A[B'-i\bar\chi(u^2,v^2)]}$ we obtain the following equation
\beq
\label{eq:saddle3}
 \frac{z^2-z_0^2}{\xi}=f_0(z)+f_u(z)u^2+f_v(z)v^2
\eeq
where $f_i(z)=2F_i'(z)/z$, with $i=0,u,v$; Explicitly:
\bea
 f_0(z)&=&\frac{1}{z^2}\left(\frac{z}{\tanh (z)}-1\right),\\
 f_u(z)&=&\frac{1}{2\cosh^2(z/2)}\left(\frac{\sinh(z)}{z}+1\right),\\
 f_v(z)&=&\frac{1}{2\sinh^2(z/2)}\left(\frac{\sinh(z)}{z}-1\right).
\eea
In terms of the solution $z\equiv z(u^2,v^2)$ of \Eqn{eq:saddle3}, we have
\beq
\label{eq:terms}
 \langle\varphi_2|D|\varphi_1\rangle= A^{-{N\over2}}{\cal D}(u^2,v^2)e^{-2iNCw} ,
\eeq
where ${\cal D}(u^2,v^2)$ is a real function given by
\beq
\label{eq:DD}
 {\cal D}(u^2,v^2)= \frac{1}{Z}e^{-N\left\{F_0(z)+F_u(z)u^2+F_v(z)v^2-\frac{\left(z^2-z_0^2\right)^2}{8\xi}\right\}}.
\eeq

We seek solutions of \Eqn{eq:saddle3} with $z^2$ real, i.e. $z$ can be either real or purely imaginary. We show in Appendix \ref{appS} that physically allowed solutions are such that $z^2>-\pi^2$. It is rather simple to see, e.g. graphically, that \Eqn{eq:saddle3} always admits one and only one such solution. Moreover, we readily see from 
the small $z$ behavior of the RHS of \Eqn{eq:saddle3}, namely $f_0(0)=f_v(0)=1/3$ and $f_u(0)=1$, that \Eqn{eq:saddle3} admits a real solution ($z^2\ge0$) if 
\beq
\label{eq:condition}
 \frac{1}{3}+u^2+\frac{v^2}{3}\ge-\frac{z_0^2}{\xi}
\eeq
and a purely imaginary one ($-\pi^2<z^2<0$) otherwise. A condition for the existence of imaginary solutions is, therefore, $z_0^2/\xi<-1/3$.
Noticing that
\beq
 \frac{z_0^2}{\xi}=\frac{1}{\kappa}\left(\frac{1}{2x}-1\right)
\eeq
this condition becomes
\beq
\label{eq:cond2}
 \kappa\le3\quad{\rm and}\quad x\ge\frac{1}{2(1-\kappa/3)}.
\eeq
In particular, we see that no imaginary solution can appear, whatever the strength of the nongaussiantity $x$ if $\kappa\ge3$, which is equivalent to $n\le n_c\approx e^{-3}$. We also see that, for $n>n_c$, the appearance of imaginary solutions is only possible for large enough nongaussianity $x$. In particular, this necessitates that $z_0^2/\xi$ be sufficiently negative and is, therefore, a nonperturbative effect.

\begin{figure}[t!]
 \epsfig{file=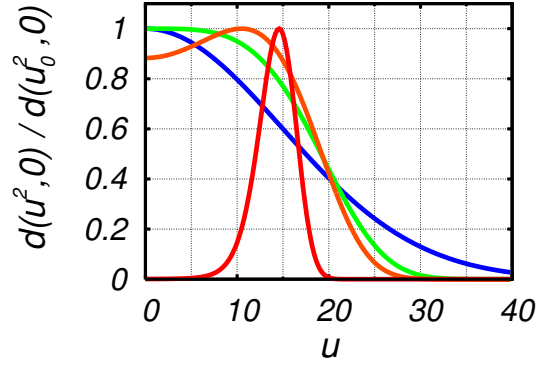,width=7.5cm}
 \caption{\label{fig:camel2} 
 The function $ d(u^2,0)$ as a function of $u$ for fixed Gaussian correlators with $n=10$ and increasing nongaussiantiy $x$: from the Gaussian case $x=0$ (blue) to $x=0.5$ (green), $x=1$ (orange) and $x=15$ (red). The corresponding values of the ratio $|C_4/2F^2|$ can be obtained from \Eqn{eq:ninf}: $0$, $0.5$, $0.67$ and $0.96$ respectively. The red curve is the $v=0$ slice of \Fig{fig:camel}. }
\end{figure}
When conditions \eqn{eq:cond2} are satisfied, there is a region of the $(u,v)$ plane -- for sufficiently small $u^2$ and $v^2$ -- where $z=z(u^2,v^2)$ is purely imaginary. Writing
\beq
 {\cal D}(u^2,v^2)=d^N\!(u^2,v^2)
\eeq
and noticing that
\beq
\label{eq:du}
 \partial_{u^2}\!\ln d(u^2,v^2) =-F_u(z).
\eeq
and 
\beq
\label{eq:dv}
 \partial_{v^2}\!\ln d(u^2,v^2) =-F_v(z)
\eeq
and that, for $z^2>-\pi^2$, ${\rm sign}(F_u(z))={\rm sign}(z^2)$ and $F_v(z)>0$, we conclude -- see Appendix \ref{app:dpeak} -- that $\ln{\cal D}(u^2,v^2)$ is always a monotonously decreasing function of $v$ whereas it either is monotonously decreasing in the $u$-direction if condition \eqn{eq:condition} is not satisfied or if $v$ is large enough: $v^2\le-z_0^2/\xi-1/3$, or has a maximum at $u=u_c(v^2)$ otherwise, corresponding to the condition  $z(u_c^2,v^2)=0$, i.e.
\beq
\label{eq:maxi}
 u_c^2(v^2)=-\frac{z_0^2}{\xi}-\frac{1}{3}-\frac{v^2}{3}.
\eeq

This is illustrated on \Fig{fig:camel}, where we plot the function $d(u^2,v^2)$ as a function of $u$ and $v$ for Gaussian parameter $n=10$ and non-Gaussian one $x=15$. We see that whenever condition \eqn{eq:cond2} is fulfilled, the non-Gaussian distribution qualitatively differs from the corresponding Gaussian one. We show, in \Fig{fig:camel2} the $v=0$ slice of the function $d$, that is essentially the probability distribution of field values $\langle\varphi|D|\varphi\rangle$, for fixed Gaussian parameter $n=10$ and various nongaussianties $x$, ranging from $x=0$ (Gaussian) to $x=15$ \footnote{We show in Appendix \ref{app:dpeak} that the width of the function $d$ in the $v$-direction is essentially independent of $x$.}. We mention that similar shapes have been also obtained in Ref. \cite{Koksma:2010zi}.

As expected, the structure of the density operators in field-configuration space inferred by either the Gaussian or the non-Gaussian observers significantly differ for large enough nongaussiantity. In the next section we further illustrate this difference by computing the Wigner distribution on phase-space and studying the quantum coherence properties inferred by both observers.

\section{Wigner function and quantum coherence}
\label{sec:Wigner}

For a Gaussian density operator the Wigner function (see below) is positive definite and can be interpreted as a probability distribution on phase-space. This remains true for the non-Gaussian density operator \eqn{eq:dmatrix} in the large-$N$ limit. One interesting property one can directly read on the Wigner function is the degree of quantum coherence of the system in the coherent-state basis.
Indeed, if the Heisenberg uncertainty principle guarantees that the area of phase space where the Wigner function is non-vanishing cannot be smaller than $1$ (in appropriate units), one can stil have a very squeezed distribution in a given direction. As recalled in Appendix \ref{appsec:coh} this gives rise to non-trivial quantum coherence between distant semi-classical (coherent) states \footnote{The degree of quantum coherence, i.e. the size of non-diagonal elements of the density matrix, of course depends on a choice of basis. Our choice to consider the coherent state basis is arbitrary here. In general the relevant basis to discuss quantum coherence depends on the particular observables one wishes to measure.}. We analyze the shape of the Wigner function for our non-Gaussian density operator in this section.

\subsection{Wigner function}

\begin{figure}[t!]
 \epsfig{file=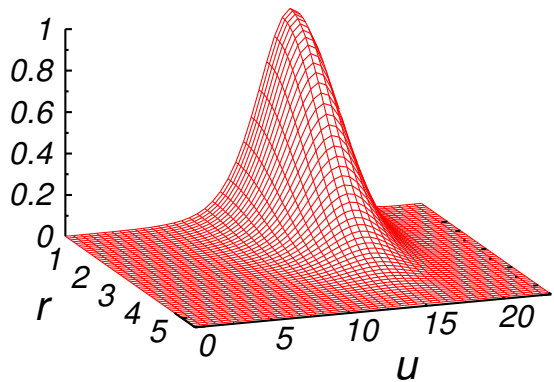,width=10.cm}
 \caption{\label{fig:Wuv} 
 The wigner function $w(u^2,r^2)$, normalized to its maximum, for the same values of parameters as in \Fig{fig:camel}.}
\end{figure}

The Wigner function is defined as
\beq
\label{eq:wigner}
 W(\vec\varphi,\vec\pi)=\int d^N\!s\,e^{i\vec\pi\cdot \vec s}\left<\varphi+s/2\right|D\left|\varphi-s/2\right>,
\eeq
where
\beq
\label{eq:wdep}
 \left< \varphi +s/2\right|D\left| \varphi-s/2\right>=A^{-{N\over2}}{\cal D}(u^2,v^2)e^{-i\frac{C}{A}\vec\varphi\cdot\vec s},
\eeq
with ${\cal D}(u^2,v^2)$ real. \Eqn{eq:wdep} and $O(N)$ symmetry imply that the Wigner function really depends on two-variables only: 
\beq
 W(\vec\varphi,\vec\pi)\equiv{\cal W}(u^2,r^2),
\eeq 
where we defined
\beq
\label{eq:r2}
 r^2=\frac{4A}{N}\left[\vec\pi-{C\over A}\vec\varphi\,\right]^2.
\eeq
Exploiting the $O(N)$ symmetry, $N-2$ angular integrations can be trivially performed in \Eqn{eq:wigner} and the remaining angular integral is given by a Bessel function $J_\nu(x)$. One obtains
\beq
\label{eq:WW}
 {\cal W}(u^2,r^2)=\frac{Nr}{2}\!\left(\frac{8\pi}{r}\right)^{\!\!{N\over2}}\!\!\int_0^\infty \!\!dv \,v^{N\over2}J_{{N\over2}-1}(Nrv)\,{\cal D}(u^2,v^2).
\eeq
It is easy to check that, at large $N$, the integral is dominated by values $v\sim1$ due to the combination of the phase-space factor $v^{N/2}$ and of the rapid decrease of the function ${\cal D}$ at large $v$. It follows that for the typical values of $r$ we are interested in, that is $r\sim1$, the argument of the Bessel function is of the same order as its index $\sim N$, which is precisely the region where the Bessel function cannot be given a simple approximation. We compute the 
large-$N$ limit of the integral \eqn{eq:WW} numerically \footnote{Specifically, for each point $(u,v)$, we compute the value of $\ln w(u^2,r^2)$ as a function of $N$ and extract the asymptotic large-$N$ limit, which is typically reached for $N\sim 10-20$.}. \Fig{fig:Wuv} shows the function $w(u^2,r^2)$ defined as 
\beq
 {\cal W}(u^2,v^2)=w^N\!(u^2,v^2)
\eeq
as a function of $u$ and $r$ for the same values of parameters as in \Fig{fig:camel}.

As for the function $d(u^2,v^2)$ before, the shape of $w(u^2,r^2)$ only depends on $n$ and $x$. This encodes the Wigner distribution on the $N^2$-dimensional phase-space, the translation to which depends on the actual values of the Gaussian correlators $F$ and $R$ through Eqs. \eqn{eq:parameters}, \eqn{eq:uvw} and \eqn{eq:r2}. Here we illustrate the content of the function $w(u^2,r^2)$ in the cases where the $O(N)$ vectors $\vec\varphi$ and $\vec\pi$ are either collinear or perpendicular to each other. It is useful to use the parametrization of Gaussian correlators introduced in \cite{Giraud:2009tn} (up to a redefinition of the angle $\phi\to\pi-\phi$):
\bea 
 F&=&\bar a\,(1+\gamma\cos\phi),\nn
 \label{eq:param}
 K&=&\bar a\,(1-\gamma\cos\phi),\\
 R&=&\bar a\,\gamma\sin\phi,\nonumber
\eea
with $\bar a=(n+1/2)/\sqrt{1-\gamma^2}$, $0\le\gamma<1$ and $0\le\phi<2\pi$. The parameter $\gamma$ controls the overall squeezing of the Wigner distribution. Note that $C/A=R/F=\tan(\phi/2)$. 

When $\vec\varphi$ and $\vec\pi$ are collinear, one has $r^2\propto (\pi-\varphi\tan(\phi/2))^2$. The angle $\phi$ gives a positive $\varphi ^2$ term, proportional to $F$, which adds to the already present one and thus squeezes the distribution in the $\varphi $-direction. It also contributes a $\varphi \pi $-term which rotates the main axis of the distribution by an angle $\phi/2$, as can be seen on \Fig{fig:proj}. In contrast, in the case where $\vec\varphi$ and $\vec\pi$ are perpendicular, one has $r^2\propto \pi ^2+ \varphi ^2\tan^2(\phi/2)$, which results in a squeezing of the distribution in the $\varphi $-direction (by the same amount as before), but there is no $\varphi \pi $-term and thus no rotation. The case $\phi=0$ corresponds to a overall squeezing (for $\gamma\neq0$) in the $\pi $-direction and an overal stretching in the $\varphi $-direction, i.e. $F>K$, whereas the case $\phi=\pi$ gives the opposite effect.

We already see from \Fig{fig:proj} that the Wigner distribution can be strongly squeezed in some direction and therefore present a high degree of quantum coherence. We discuss this further in the next section and compare to the Gaussian case.

\subsection{Quantum coherence}

\begin{figure}[t!]
 \epsfig{file=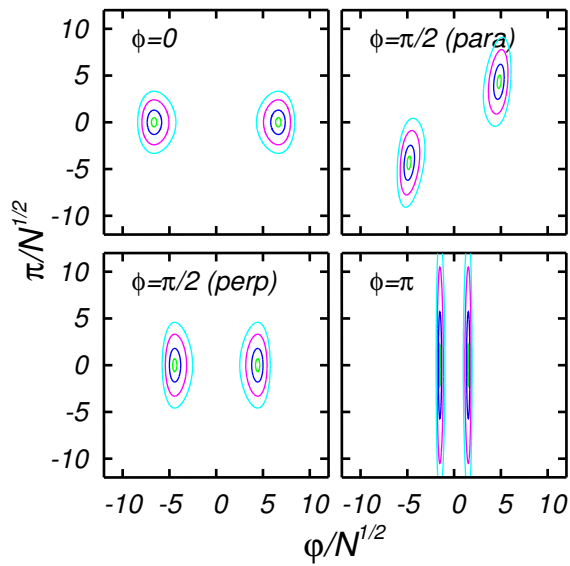,width=9.5cm}
 \caption{\label{fig:proj} 
 Contour plots of the Wigner distribution $w$ in physical coordinates for the cases where $\vec\varphi$ and $\vec\pi$ are either parallel ($para$) or perpendicular ($perp$) in $O(N)$ space, for the same parameters as in \Fig{fig:Wuv}. We use the parametrization \eqn{eq:param} with $n=10$ and  $\gamma=0.9$. It should be kept in mind that the above figures are two-dimensional slices of a $O(N)$-symmetric distribution in $N^2$-dimensional phase-space. In particular, the two apparently separated structures in each figure are in fact connected by $O(N)$-transformations.}
\end{figure}

To illustrate the implications of the very different Gaussian vs. non-Gaussian Wigner functions for quantum coherence, we consider the case where the correlator $R=0$, that is $r^2=4A \pi ^2/N$. As described before, for sufficiently large values of $n$ and $x$ \footnote{One has to keep in mind that the value of $x$ is actually fixed by the value of the ``measured'' non-Gaussian correlator $C_4/2F^2$. Distinguishing between various values of $x$ at large $x$ requires a precise knowledge of the latter, see e.g. \Eqn{eq:ninf}. It is assumed throughout this paper that the measured correlators $F$, $K$, $R$ and $C_4$ are known with infinite precision.}, the function $d(u^2,v^2)$ exhibits a clear peak-structure in the $(u,v)$-plane, located at $u=u_0$, $v=0$, with, see \Eqn{eq:maxi}, $u_0^2=-z_0^2/\xi-1/3$. We find that for $x\gg n^2\gg1$ the peak is well-described by a Gaussian:
\beq
 d(u^2,v^2)\approx d(u^2_0,0)e^{-\frac{(u-u_0)^2}{2\delta_u^2}-\frac{v^2}{2\delta_v^2}},
\eeq
with, see Appendix \ref{app:dpeak}, 
\beq
 u_0^2=2n^2,\quad\delta_u^2={1\over6}+\frac{n^2}{2x}\quad{\rm and}\quad \delta_v^2={1\over2}.
\eeq
In that case, the Wigner function is also Gaussian:
\beq
\label{eq:wiggauss}
 w(u^2,r^2)\approx w(u^2_0,0)e^{-\frac{(u-u_0)^2}{2\delta_u^2}-\frac{r^2}{2\delta_r^2}},
\eeq
with
\beq
 \delta_r^2=2.
\eeq
In terms of physical variables $\varphi$ and $\pi $, the distribution peaks at $\varphi_0^2/N=Au_0^2$, with widths $\delta_ \varphi ^2/N=A\delta_u^2$ and $\delta_ \pi ^2/N=\delta_r^2/4A$, where $A=\kappa F\approx F/2n^2\approx1/2K$, that is:
\beq
\label{eq:physical}
 {\varphi_0^2\over N}=F,\qquad\frac{\delta_ \varphi ^2}{N}=\frac{F}{12n^2}\quad{\rm and}\quad \frac{\delta_ \pi ^2}{N}=K.
\eeq
Note that the width in the $\pi $-direction is unchanged compared to the corresponding Gaussian ($x=0$) case.  In contrast, the width in the $\varphi $-direction is much smaller than the corresponding Gaussian one $\sim F$ \footnote{Note that the product 
 ${\delta_q\delta_ \pi/N}= {\delta_u/\sqrt 2}>{1/\sqrt{12}}$ is bounded from below. We emphasize that the small bound is not in contradiction with the Heisenberg inequality. The latter would hold for well-separated (disconnected) parts of the Wigner distribution in phase space, which is not the case here due to the underlying $O(N)$ symmetry.}. 

\begin{figure}[t!]
 \epsfig{file=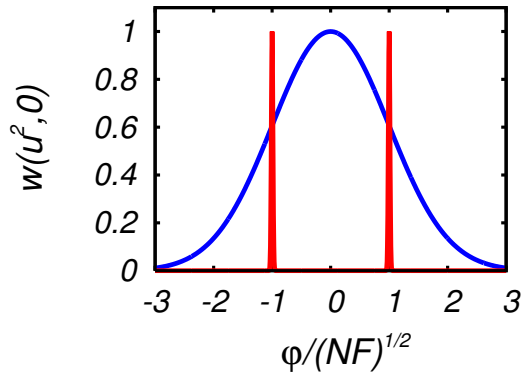,width=7.5cm}
 \caption{\label{fig:illu} 
 The function $w(u^2,0)$ normalized to its maximum value plotted against the physical coordinate normalized to the overall width of the distribution: $\varphi/\sqrt{NF}$, . The blue and red curves are the distribution inferred by the Gaussian and the non-Gaussian observers respectively for $n=10$ (in the limit $x\gg n^2$ in the latter case). While the former sees a broad, i.e. incoherent, distribution, the latter instead finds highly squeezed distribution aroun $\varphi^2=\varphi_0^2$ and, therefore, a highly coherent state. As already emphasized in \Fig{fig:proj}, the two peaks are in fact connected by the underlying $O(N)$ symmetry and should not be interpreted as an incoherent superposition of two states.}
\end{figure}

We see that our observers may arrive at very different conclusions concerning the degree of quantum coherence of the system. Indeed, suppose that $F\sim K\sim n$, in which case the Gaussian observer concludes to a statistical mixture of many uncorrelated semi-classical states, see e.g. \cite{Campo:2008ju,Giraud:2009tn}. In constrast, the non-Gaussian observer gets an $O(N)$-symmetric, highly squeezed Wigner distribution around a non-zero value $\varphi=\varphi_0$, with $\delta_ \varphi/\sqrt N\sim1/n\ll1$ and thus concludes to a high degree of quantum coherence (see Appendix \ref{appsec:coh}). This situation is illustrated in \Fig{fig:illu}.

Also interesting is the case where $F\sim 1$ and $K\sim n^2$, for which the Gaussian observer concludes to what Campo and Parentani call a partially decohered distribution \cite{Campo:2008ju}, but where the non-Gaussian observer will get, again, a highly squeezed Wigner distribution, with $\delta_ \varphi\sim 1/n^2\ll1$, thus exhibiting a high degree of quantum coherence. This situation can be observed for instance in the right-bottom ($\phi=\pi$) plot of \Fig{fig:proj}, which shows a slice in $N^2$-dimensional phase-space. Here the corresponding Gaussian distribution would be a long ellipse in the $\pi$-direction surrounding the two highly squeezed contour plots corresponding to the non-Gaussian distribution.

\section{Conclusion}

In this study we compare the effective density operators inferred from two observers, one of which only measures Gaussian correlators $F$, $K$ and $R$ and the other has further access to the field four-point function $C_4$. As can be expected, we find that, depending on the strength of the measured nongaussiantity (quantified by our parameter $x$), both observers can arrive at very similar or very different conclusions concerning various observables. 

Global observables such as the correlation entropy or the quantum purity are rather insensitive to the detailed structure of the underlying density matrix and thus to the amount of nongaussianity. In other words, both observers essentially agree about the degree of quantum purity of the system. 

In contrast, more detailed observables such as the probability distribution of field configurations, can be very different for sufficiently large nongaussianity. We identify situations where the Gaussian observer would conclude to a thermal-like, incoherent mixture of a large number of states whereas the extra piece of information the non-Gaussian observer has leads him to conclude instead to a state with a high degree of quantum coherence.

An interesting extension of the present study will be to include other possible four-point correlators in the non-Gaussian description,  such as $\langle\pi^4\rangle$ or $\langle\pi^2\varphi^2\rangle$. In particular, it will be interesting to see whether these change the present conclusions concerning entropy or purity.

In the context of recent studies of entropy production and decoherence in quantum field theory based on the incomplete descritpion picture \cite{Giraud:2009tn,Koksma:2009wa}, the present work provides a basis to study the role of dynamically generated non-Gaussian correlators concerning entropy production or the decoherence process. 

Finally, the present work is also of relevance for more general studies of nonequilibrium dynamics and thermalization in quantum field theory \cite{Berges:2004vw}. The existing literature, see e.g. \cite{Berges:2001fi,Berges:2002wr, Cooper:2002qd, Juchem:2003bi, Arrizabalaga:2005tf, Calzetta:2003dk, Tranberg:2008ae} is, to its vast majority based on analyzing thermalization from a Gaussian observer perspective, i.e. looking at the way equal-time two-point functions approach their equilibrium values \footnote{An exception is Ref. \cite{Berges:2002wr}, where information from the nonequal-time two-point functions is used to study the onset of the thermal fluctuation-dissipation relation.}. We believe our work opens the way for more detailed studies of the nonequilibrium flow in the space of correlation functions.

\section*{Acknowledgements}
We thank R. Parentani for interesting comments as well as J. Koksma, T. Prokopec and M. G. Schmidt for useful information concerning Ref. \cite{Koksma:2010zi}.

\appendix
\section{Connected four-point correlator $C_4$ at large-$N$} 
\label{appFT}

We derive the expression of the connected correlator $C_4$ in the limit $N\to\infty$. The latter can be obtained in various standard ways \cite{ZJ}, for instance by computing $\ln Z$ at next-to-leading order in the $1/N$-expansion and using \eqn{eq:deta}, or by introducing a linearly coupled source term $j_a\varphi_a$ in the classical action and taking four functional derivatives of the resulting generating functional with respect to $j$, see e.g. \cite{Cooper:1994hr}. We present here an alternative, simple derivation, based on introducing a bilinearly coupled source $J\varphi_a\varphi_a$.
Consider the following functional (note that $Z[J=0]=Z$)
\beq
\label{appeq:ZJ}
 Z[J]=\int_{\rm per.}{\cal D}\varphi{\cal D}\pi\,e^{\int_0^1 d\tau\,\left\{i\dot\varphi_a\pi_a-{\cal F}(\varphi,\pi)+J\varphi_a\varphi_a\right\}}.
\eeq
Using similar manipulations as before, it is an easy exercice to check that, up to an irrelevant constant,
\beq
\label{appeq:ZJ2}
 \hspace{-.3cm}Z[J]\propto{\cal N}\left(\frac{\eta}{N}\right)\int\!\!{\cal D}\chi\,e^{-\frac{N}{2}\left({\rm Tr}\,{\rm Ln}G^{-1}(\chi)+\int_0^1\!\!d\tau\frac{(\chi+iJ)^2}{2\eta}\right)}.
\eeq
Computing $\delta\ln Z[J]/\delta J(\tau)|_{J=0}$ from either \Eqn{appeq:ZJ} or \Eqn{appeq:ZJ2}, one gets the exact relation
\beq
\label{appeq:chi}
 \langle\chi\rangle=2i\eta F,
\eeq
where 
\beq
 \langle F[\chi]\rangle=\frac{\int\!\!{\cal D}\chi\,F[\chi]\,e^{-\frac{N}{2}\left({\rm Tr}\,{\rm Ln}G^{-1}(\chi)+\int_0^1\!\!d\tau\chi^2/2\eta\right)}}{\int\!\!{\cal D}\chi\,e^{-\frac{N}{2}\left({\rm Tr}\,{\rm Ln}G^{-1}(\chi)+\int_0^1\!\!d\tau\chi^2/2\eta\right)}}.
\eeq
Similarly, expressing $\partial\ln Z[0]/\partial B$, one shows that
\beq
 F=2A\langle{\rm Tr} G(\chi)\rangle.
\eeq
In particular, one has the exact relation
\beq
 \langle\chi\rangle=4iA\eta \langle{\rm Tr} G(\chi)\rangle,
\eeq
of which \Eqn{eq:gap} is the expression in the limit $N\to\infty$.

Next, from \Eqn{appeq:ZJ2}, we get \footnote{We use ${\rm Tr} 1\equiv\int_0^1 d\tau \delta(\tau-\tau)=\delta(0)$.}
\beq
 \partial_\eta\ln Z=-\frac{\delta(0)}{2\eta}+\frac{N}{4\eta^2}\int_0^1 d\tau\,\langle\chi^2(\tau)\rangle.
\eeq
Writing
\beq
 \langle\chi(\tau)\chi(\tau')\rangle=-(2\eta F)^2+\langle\chi(\tau)\chi(\tau')\rangle_{_c},
\eeq
where $\langle\ldots\rangle_{_c}$ denotes the connected contribution and where we used \eqn{appeq:chi}, and using \Eqn{eq:deta}, we obtain the exact relation
\beq
\label{appeq:exact}
 2F^2+\frac{N+2}{N}C_4=\frac{\delta(0)}{2\eta}-\frac{N}{4\eta^2}\int_0^1 d\tau\,\langle\chi^2(\tau)\rangle_c.
\eeq

Finally, we note that the connected $\chi$-correlator can be obtained from
\beq
\label{appeq:correl}
 \left.\frac{1}{N}\frac{\delta^2\ln Z[J]}{\delta J(\tau)\delta J(\tau')}\right|_{J=0}=\frac{1}{2\eta}\left(\delta(\tau-\tau')-\frac{N}{2\eta}\langle\chi(\tau)\chi(\tau')\rangle_c\right).
\eeq

We now consider the large-$N$ limit. Up to an irrelevant additive constant,
\beq
\label{appeq:ZJN}
 \frac{\ln Z[J]}{N}=-\frac{1}{2}{\rm Tr}\,{\rm Ln}G^{-1}(\bar\chi_J)-\frac{(\bar\chi_J+iJ)^2}{4\eta},
\eeq
where $\bar\chi_J$ satisfies the saddle-point equation
\beq
 \bar\chi_J+iJ=4iA\eta{\rm Tr}G(\bar\chi_J).
\eeq
Note that $\bar\chi_{J=0}=\bar\chi$, see \Eqn{eq:gap}.
Differentiating both sides with respect to $J$, one easily shows that
\beq
 \left.\frac{\delta\bar\chi_J(\tau)}{\delta J(\tau')}\right|_{J=0}=-iD(\tau-\tau'),
\eeq
where the function $D$ is defined by ($\mathds{1}\equiv \delta(\tau-\tau')$)
\beq
\label{appeq:dpi}
 D^{-1}=\mathds{1}+\eta\,\Pi\quad{\rm with}\quad\Pi=16A^2G^2(\bar\chi).
\eeq
In particular, one has
\bea
 D&=&\mathds{1}-\eta\,\Pi\star D\nn
 &=&\mathds{1}-\eta\,\Pi+\eta^2\,\Pi\star\Pi\star D,
\eea
where $[A\star B](\tau,\tau')\equiv\int d\tau''\,A(\tau,\tau'')B(\tau''\tau')$.
Differentiating \Eqn{appeq:ZJN} twice with respect to $J$ and using \Eqn{appeq:correl}, one obtains the familiar expression of the $\chi$-field correlator in the large-$N$ limit \cite{ZJ}:
\beq
 \langle\chi(\tau)\chi(\tau')\rangle_{_c}=\frac{2\eta}{N}D(\tau-\tau').
\eeq
Inserting the latter in \Eqn{appeq:exact} and using ${\rm Tr} \,\Pi=G^2(\bar\chi;0)=[{\rm Tr}\,G(\bar\chi)]^2=(2F)^2$, we get:
\beq
\label{appeq:c4}
 C_4=-\frac{\eta}{2}\,{\rm Tr}\,[\Pi\star\Pi\star D].
\eeq

\section{Matsubara sums}
\label{appFT1}

Here, we compute the functional traces needed in Eqs. \eqn{eq:gap} and \eqn{appeq:c4}, using standard methods of finite temperature field theory \cite{Lebellac}. The basic quantity is the Green's function \eqn{eq:green} evaluated for a $\tau$-independent field configuration $\chi(\tau)=\bar\chi$. Writing 
\beq
 G(\bar\chi;\tau-\tau')\equiv g(\tau-\tau',z),
\eeq
where $z=2\sqrt{A(B'-i\bar\chi)}$, we have
\beq
 -g''(\tau,z)+z^2g(\tau,z)=\delta(\tau).
\eeq
Periodic boundary conditions of the functional integral \eqn{eq:funcint} results in the periodicity of the function $g$: $g(\tau+1)=g(\tau)$ which can thus be written as a Fourier series:
\beq
\label{appeq:matsu}
 g(\tau,z)=\sum_{n\in \mathds{Z}}\,e^{2i\pi n\tau}g_n(z),
\eeq
where
\beq
 g_n(z)=(\omega_n^2+z^2)^{-1}\,,\quad\omega_n=2\pi n.
\eeq
Inverting \eqn{appeq:matsu}, one obtains
\beq
 g(\tau,z)=\frac{1}{2z}\left([1+n(z)]e^{-z\tau}+n(z)e^{z\tau}\right),
\eeq
where $n(z)=(e^z-1)^{-1}$. \Eqn{eq:gap2} follows directly from
\beq
 {\rm Tr}\, G(\bar\chi)=g(0,z)=\sum_{n\in \mathds{Z}}g_n(z)=\frac{1}{2z\tanh(z/2)}.
\eeq

The function $\Pi=16A^2G^2(\bar\chi)$, see \Eqn{appeq:dpi}, is easily obtained. Using the relations
\beq
 n(z+z')[1+n(z)+n(z')]=n(z)n(z')
\eeq
and
\beq 
 n'(z)=-n(z)[1+n(z)]
\eeq
one sees that the $\tau$-dependence of the square of the Green's function $g(\tau,z)$ with ``frequency'' $z$ is given by that of the same Green's function with frequency $2z$:
\beq
 g^2(\tau,z)=2g(0,z)g(\tau,2z)-\frac{n'(z)}{2z^2}.
\eeq
Introducing
\beq
 \theta_n(z)=32A^2[g(0,z)-n'(z)\delta_{n,0}]
\eeq
the Fourier components of the function $\Pi$ read 
\beq
 \Pi_n(z)=\theta_n(z)g_n(2z).
\eeq

Finally, we write the Fourier coefficients of the $\chi$-propagator $D$ in \Eqn{appeq:dpi} as
\beq
 D^{-1}_n(z)=1+\eta\,\Pi_n(z).
\eeq
One easily checks that the $\tau$-dependence of $D$ is essentially given by that of the Green's function $g$ with a modified ``frequency''. Specifically, one has:
\beq
 \Pi_n(z)D_n(z)=\theta_n(z)g_n\left(\sqrt{(2z)^2+\eta\,\theta_n(z)}\right).
\eeq

The fact that, for what concerns their $\tau$-dependence, both $\Pi$ and $\Pi\star D$ are essentially given by the Green's function $g$ allows us to rewrite the double-convolution  in the expression \eqn{appeq:c4} of $C_4$ as a simple one-loop--like expression. For $z$ solution of the saddle-point equation \eqn{eq:gap2}, one has $n(z)=n$, i.e. $z=\ln(1+1/n)$ and $g(0,z)=1/2\kappa$. We obtain
\beq
 \frac{C_4}{\left(16\kappa F^2\right)^2}=-\frac{\eta}{2}\left\{\sum_{n\neq0}g_n(2z)g_n(2z')+\zeta^2g_0(2z)g_0(2z'')\!\right\}\!,
\eeq
where $z'=z\sqrt{1+x}$ and $z''=z\sqrt{1+\zeta x}$. The sum is readily performed using \cite{Lebellac}
\beq
 \sum_{n\in \mathds{Z}}g_n(x)g_n(x')=\frac{1}{2xx'}\left\{\frac{h(x)+h(x')}{x+x'}-\frac{h(x)-h(x')}{x-x'}\right\},
\eeq
where $h(x)=n(x)+1/2$. Equation \eqn{eq:c4} follows, with $yf(y)=h(2zy)$.

\section{On solutions of saddle-point equations}
\label{appS}

The evaluation of both $\ln Z$ and $\langle\varphi_2|D|\varphi_1\rangle$ in the limit $N\to\infty$ involve Gaussian functional integrals of the form 
\beq
 \int {\cal D}\varphi\,e^{-{\cal S}[\varphi]}
\eeq
with different types of boundary conditions, where 
\beq
 {\cal S}[\varphi]=\frac{1}{4A}\int_0^1 d\tau\,\varphi\left(-\frac{d^2}{d\tau^2}+z^2\right)\varphi,
\eeq
with $z$ the solution of the relevant saddle-point equation. We discuss the physically allowed values of $z$, i.e. those for which these integrals are well-defined. We consider only one $O(N)$ component for the sake of the argument. 

The calculation of $\ln Z$ involves an integral over fields with periodic boundary conditions $\varphi(0)=\varphi(1)$ whereas $\langle\varphi_2|D|\varphi_1\rangle$ involves fixed boundary conditions $\varphi(0)=\varphi_1$ and $\varphi(1)=\varphi_2$. In the latter case, one can compute the explicit dependence in $\varphi_1$ and $\varphi_2$:
\beq
\label{appeq:neuman}
 \int_{\varphi_1}^{\varphi_2} {\cal D}\varphi\,e^{-{\cal S}[\varphi]}=e^{-F_u(z)u^2-F_v(z)v^2}\int_0 {\cal D}\varphi\,e^{-{\cal S}[\varphi]},
\eeq
where $u^2=(\varphi_2+\varphi_1)^2/4A$, $v^2=(\varphi_2-\varphi_1)^2/4A$ and the functions $F_u(z)$ and $F_v(z)$ are given in Eqs. \eqn{eq:fu} and \eqn{eq:fv}. The functional integral on the RHS is to be evaluated with Neuman boundary conditions: $\varphi(0)=\varphi(1)=0$.

Writing $\varphi(\tau)=\sum_n e^{i\omega_n\tau}\varphi_n$, one is finally led to evaluate Gaussian functional integrals of the form
\beq
\label{appeq:fff}
 \int {\cal D}\varphi\,e^{-\frac{1}{4A}\sum_n|\varphi_n|^2(\omega_n^2+z^2)},
\eeq
with different types of boundary conditions. Those are well-defined if $\omega_n^2+z^2>0, \forall n$.

For Neuman boundary conditions $\varphi(0)=\varphi(1)=0$, one has $\omega_n=n\pi$ with $n\in\mathds{N}^\star$. The lowest frequency in \eqn{appeq:fff} is $\omega_1=\pi$ and the integral is well-defined for $z^2>-\pi^2$.

For periodic boundary conditions $\varphi(0)=\varphi(1)$, the frequencies $\omega_n=2\pi n$ with $n\in\mathds{Z}$ and the lowest one is $\omega_0=0$. The functional integral is well defined only for $z^2>0$. Alternatively, the integral with periodic boundary conditions can be obtained from the one with fixed boundary conditions as
\bea
 \int_{\rm per.} {\cal D}\varphi\,e^{-{\cal S}[\varphi]}&=&\int d\tilde\varphi\int_{\tilde\varphi}^{\tilde\varphi} {\cal D}\varphi\,e^{-{\cal S}[\varphi]}\nn
 &=&\int_0 {\cal D}\varphi\,e^{-{\cal S}[\varphi]}\int d\tilde\varphi\,e^{-F_u(z)\frac{\tilde\varphi^2}{A}}\!,\quad
\eea
where we used \eqn{appeq:neuman}. For real $z^2$, the $\tilde\varphi$-integral is well-defined if $F_u(z)>0$, which is equivalent to $z^2>0$.

\section{Wigner function and quantum coherence}
\label{appsec:coh}

Here we briefly recall the relation between the shape of the Wigner distribution on phase-space and the degree of quantum coherence between distinct elementary cells of phase-space, described by coherent states. The latter are the eigenstates of the usual annihilation operator \footnote{We stick to a one-dimensional degree of freedom, corresponding to the zero Fourrier mode of the quantum scalar field. For a non-zero Fourrier mode, one needs to consider two-modes coherent states, see e.g. \cite{Campo:2008ju}.}: $a_a|\alpha\rangle=\alpha_a|\alpha\rangle\,, \forall a$, with
\beq
 a_a=\frac{\hat\varphi_a+i\hat\pi_a}{\sqrt 2}.
\eeq

Recall that the Wigner distribution corresponding to a given coherent state is a Gaussian centered in $\sqrt2({\rm Re }\alpha_a,{\rm Im}\alpha_a)$ and of width $1/\sqrt 2$ in both $\varphi$ and $\pi$ directions, in units of the natural field dimension \cite{Glauber:1963tx, Campo:2008ju}. Because in that state the quantum fluctuations in both $\varphi$ and $\pi$ are as small as they can be according to the Heisenberg uncertainty principle, it is often called semi-classical states. It can be seen as describing an elementary quantum cell in phase-space. 

Using
\beq
 \langle\varphi|\alpha\rangle=\frac{1}{\pi^{N/4}}e^{-\frac{1}{2}\left(\vec\varphi-\frac{\vec\alpha^*+\vec\alpha}{\sqrt 2}\right)^2+\frac{\vec\alpha^*-\vec\alpha}{\sqrt 2}\cdot\vec\varphi+\frac{\vec\alpha^{*2}-\vec\alpha^2}{4}}
\eeq
and 
\beq
 \langle\alpha|\alpha'\rangle=e^{-\frac{|\vec\alpha|^2}{2}-\frac{|\vec\alpha'|^2}{2}+\vec\alpha^*\cdot\vec\alpha'}
\eeq
where $|\vec\alpha|^2\equiv\vec\alpha^*\cdot\vec\alpha$, one easily derives the following formula for the matrix elements of a given density matrix $D$ in terms of the corresponding Wigner function:
\beq
 \frac{\langle\alpha|D|\alpha'\rangle}{\langle\alpha|\alpha'\rangle}\!=\!\int\frac{d^N\! \varphi\, d^N\!\pi}{\pi^N}\,W(\vec \varphi,\vec \pi)\,e^{-\left(\vec \varphi-\vec\beta_ \varphi\right)^2-\left(\vec \pi-\vec\beta_\pi\right)^2}
\eeq
where $\vec\beta_ \varphi =\frac{\vec\alpha'+\vec\alpha^*}{\sqrt 2}$ and $\vec\beta_\pi=\frac{\vec\alpha'-\vec\alpha^*}{i\sqrt 2}$.

Let us first recall the simple case of a Gaussian Wigner function centered at the origin \footnote{The widths introduced here are related to the those defined in \Eqn{eq:physical} by $\Delta_{\varphi,\pi}^2=\delta_{\varphi,\pi}^2/N$.}: 
\beq
 W(\vec \varphi,\vec \pi)\propto  e^{-{\varphi^2\over 2\Delta_ \varphi ^2}-{\pi^2\over 2\Delta_\pi^2}},
\eeq
for which the calculation is straightforward.
In terms of real and imaginary parts $\vec \alpha=\vec a+i\vec b$ and $\vec \alpha'=\vec a'+i\vec b'$, one obtains
\beq
\label{appeq:coherence}
 \left|\langle\alpha|D|\alpha'\rangle\right|\propto e^{-\frac{(\vec a+\vec a')^2}{2(1+2\Delta_ \varphi ^2)}-\frac{(\vec b+\vec b')^2}{2(1+2\Delta_\pi^2)}}\times e^{-\frac{\Delta_\pi^2(\vec a-\vec a')^2}{1+2\Delta_\pi^2}-\frac{\Delta_ \varphi ^2(\vec b-\vec b')^2}{1+2\Delta_ \varphi ^2}}.
\eeq

One thus observes non-trivial correlations -- quantum coherence -- between distant phase-space cells  ($|\vec \alpha-\vec\alpha'|\gtrsim1$) when the Wigner distribution is sufficiently squeezed in either the $\pi$  ($\Delta_\pi\lesssim1$) or the $\varphi $ ($\Delta_ \varphi\lesssim1$) direction. Note that the size of the correlation in the $\varphi $ (resp. $\pi$) direction -- i.e. along the real (resp. imaginary) axis in the coherent state basis -- is solely controlled by the width of the Wigner distribution
in the $\pi$ (resp. $\varphi $) direction.

Consider now the case of interest in \Eqn{eq:wiggauss}: 
\beq
 W(\vec \varphi,\vec \pi)\propto  e^{-{(\varphi -\varphi_0)^2\over 2\Delta_ \varphi ^2}-{\pi^2\over 2\Delta_\pi^2}}.
\eeq
The $\pi$-integration goes as in the previous case. As for the $\varphi$-integration, $N-1$ angular integrals are easily performed and one is left with the following radial integral:
\beq
\label{appeq:radial}
 \int_0^\infty\!\!d\varphi\,\varphi^{N\over2}e^{-{(\varphi -\varphi_0)^2\over 2\Delta_ \varphi ^2}-\varphi^2}J_{{N\over2}-1}(2i\beta_\varphi \varphi),
\eeq
with $\beta_\varphi=(\vec\beta_\varphi\cdot\vec\beta_\varphi)^{1/2}$. For $N\gg1$ the factor before the Bessel function is strongly peaked around $\varphi=\varphi_*$ which, in the case of interest here, see \Eqn{eq:physical} with $F/n^2\ll1$, is given by $\varphi_*\approx\varphi_0$. Depending on the value of $|\beta_\varphi|$ in \eqn{appeq:radial}, one may employ different approximations for the Bessel function and evaluate the integral. For instance, for $|\beta_\varphi \varphi_0|\gg N/2$ one can use the leading asymptotic behavior $J_\nu(z)\sim\sqrt{2/\pi z}\,\cos(z-\nu\pi/2-\pi/4)$ and is then left with the evaluation of Gaussian integrals. We obtain finally, in terms of real and imaginary parts of $\vec\alpha$ and $\vec\alpha'$:
\bea
 &&\left|\langle\alpha|D|\alpha'\rangle\right|\propto \left(\frac{\varphi_0}{|\beta_\varphi|}\right)^{\!\!{N\over2}}|\cos(2i\beta_\varphi\varphi_0-N\pi/4)|\nn
 &&\qquad e^{-\frac{(\vec a+\vec a')^2}{2(1+2\Delta_ \varphi ^2)}-\frac{(\vec b+\vec b')^2}{2(1+2\Delta_\pi^2)}}\times e^{-\frac{\Delta_\pi^2(\vec a-\vec a')^2}{1+2\Delta_\pi^2}-\frac{\Delta_ \varphi ^2(\vec b-\vec b')^2}{1+2\Delta_ \varphi ^2}}.\nn
\eea

As an illustration, consider two phase-space cells such that $\vec \alpha=\vec\alpha^{\prime\star}=(\varphi_0\vec e+i\vec x)/\sqrt{2}$. One gets:
\beq
 \left|\langle\alpha|D|\alpha'\rangle\right|\propto \left(\frac{\varphi_0^2}{\varphi_0^2+x^2}\right)^{\!\!{N\over4}}e^{-\frac{2\Delta_ \varphi ^2}{1+2\Delta_ \varphi ^2}x^2}.
\eeq
The conclusions drawn from the previous case, \Eqn{appeq:coherence}, concerning the degree of quantum coherence still hold.

\section{Peak structure of $d(u^2,v^2)$}
\label{app:dpeak}

We discuss the structure of the matrix element \eqn{eq:terms} for $n$ and $x$ sufficiently large so that a clear peak structure appears, see e.g. \Fig{fig:camel}. For $n\gg1$ and $x\gg1$ the latter is well-described by a Gaussian. Recalling Eqs. \eqn{eq:du} and \eqn{eq:dv}, we have
\bea
 \partial_u\ln d&=&-2u\,F_u(z)\\
 \partial_v\ln d&=&-2v\,F_v(z),
\eea 
where $z\equiv z(u^2,v^2)$.
Since $F_v(z)>0$ for real $z^2$,  the only extremum of $\ln d$ in the $v$-direction is located at $v=0$. It is a maximum. Instead, $\partial_u\ln d$ vanishes both at $u=0$ and at the point where $F_u(z)=0\Leftrightarrow z(u^2,v^2)=0$, i.e. $u^2=u_c^2(v^2)$ see \Eqn{eq:maxi}. The latter extremum only exists if conditions \eqn{eq:condition} are satisfied, in which case it is a maximum whereas the one at $u=0$ is a local minimum. We consider this case in the following. 

The absolute maximum of the function $\ln d$ in the $(u,v)$-plane is therefore located at $({u_0},0)$ with 
\beq
 u_0^2\equiv u_c^2(0)=-\frac{z_0^2}{\xi}-\frac{1}{3}.
\eeq
Performing a Taylor expansion near this point (denoted by a subscript $0$, one has
\beq
 \partial_u^2\ln d|_{_0}=\left.-4u_0^2F_u^\prime(z)\partial_{u^2} z\right|_{z\to0},
\eeq
\beq
 \partial_v^2\ln d|_{_0}=-2F_v(0)
\eeq
and the cross-derivative $\partial_u\partial_v\ln d|_{_0}=0$.
From the saddle-point  equation \eqn{eq:saddle3}, one gets
\beq
 \partial_{u^2} z=\frac{f_u(z)}{\frac{2z}{\xi}-[f_0'(z)+f_u'(z)u^2+f_v'(z)v^2]}
\eeq
Using small-$z$ behaviors of the functions $F_{i}(z)$ and $f_i(z)$, with $i=0,u,v$,  we get, after calculations,
\beq
 \ln d=\ln d_0-{(u-u_0)^2\over2\delta_u^2}-{v^2\over2\delta_v^2} + \ldots
\eeq
where $d_0=d(u_0^2,0)$, with
\beq
 \delta_u^2=\frac{1}{u_0^2}\left({1\over\xi}+{u_0^2\over6}+{1\over45}\right)\quad{\rm and}\quad\delta_v^2={1\over2}.
\eeq
For  $n\gg1$, one has $u_0^2\approx 2n^2(1-1/2x)$, $\xi\approx x/n^4$ and
\beq
 \delta_u^2=\frac{n^2}{2x-1}+{1\over6}+{\cal O}\left(n^{-2}\right)\quad{\rm and}\quad \delta_v^2={1\over2}.
\eeq
The result of main text follows in the limit $x\gg n^2$.
 
Finally, we note that a similar calculation yields, for the first non-vanishing higher-order derivatives, in the same limit,
\beq
 \partial_u^3\ln d|_{_0}=\frac{2}{3u_0\delta_u^2}\left(\frac{1}{\delta_u^2}-\frac{9}{2}\right)\sim n^{-1},
\eeq
\beq
 \partial_u^2\partial_v^2\ln d|_{_0}=\frac{2}{15u_0^2\delta_u^2}\left(\frac{1}{\delta_u^2}-\frac{5}{2}\right)\sim n^{-2}
\eeq
and
\beq
 \partial_v^4\ln d|_{_0}=-\frac{1}{3u_0^2\delta_u^2}\sim n^{-2}.
\eeq
The function $d$ is thus well-described by a Gaussian, which is confirmed by our numerical results.


\begin{thebibliography}{10}

\bibitem{Balian}
 R.~Balian, Am.\ J.\ Phys. {\bf 68} (2000) 1060.
    
\bibitem{Calzetta:2003dk}
  E.~A.~Calzetta and B.~L.~Hu,
  Phys.\ Rev.\  D {\bf 68} (2003) 065027;
 See also {\it Nonequilibrium Quantum Field Theory}, Cambridge University Press (2008).

\bibitem{Campo:2008ij}
  D.~Campo, R.~Parentani,
  Phys.\ Rev.\  {\bf D78}, 065045 (2008). 
  For earlier work in a similar spirit, see also
  R.~H.~Brandenberger, T.~Prokopec and V.~F.~Mukhanov,
  Phys.\ Rev.\  D {\bf 48} (1993) 2443.

\bibitem{Campo:2008ju}
  D.~Campo, R.~Parentani,
  Phys.\ Rev.\  D {\bf 78} (2008) 065044;
 Phys.\ Rev.\  D {\bf 72} (2005) 045015.

\bibitem{Giraud:2009tn}
  A.~Giraud, J.~Serreau,
  Phys.\ Rev.\ Lett.\  {\bf 104} (2010) 230405.

\bibitem{Koksma:2009wa}
  J.~F.~Koksma, T.~Prokopec and M.~G.~Schmidt,
  Phys.\ Rev.\  D {\bf 81} (2010) 065030;
  arXiv:1012.3701 [quant-ph];
  arXiv:1101.5323 [quant-ph];
  arXiv:1102.4713 [hep-th].

\bibitem{Polarski:1995jg}
  D.~Polarski, A.A.~Starobinsky,
  Class.\ Quant.\ Grav.\  {\bf 13} (1996) 377;
  J.~Lesgourgues, D.~Polarski, A.A.~Starobinsky,
  Nucl.\ Phys.\  B {\bf 497} (1997) 479.

\bibitem{Herranen:2008di}
  M.~Herranen, K.~Kainulainen, P.M.~Rahkila,
  JHEP {\bf 0905} (2009) 119;
  Nucl.\ Phys.\  B {\bf 810} (2009) 389;
  JHEP {\bf 1012} (2010) 072.

\bibitem{Giunti:2002xg}
  C.~Giunti,
  JHEP {\bf 0211} (2002) 017;
  G.G.~Raffelt, G.~Sigl,
  Phys.\ Rev.\  D {\bf 75} (2007) 083002;

\bibitem{Muller:2005yu}
  B.~M\"uller, A.~Sch\"afer,
  Phys.\ Rev.\  C {\bf 73} (2006) 054905.

\bibitem{Graham}
 R.~Graham,
 Phys.\ Rev.\ Lett.\ {\bf 81} (1998) 5262.
 
\bibitem{Berges:2004vw}
  For short reviews, see J.~Berges, J.~Serreau, {\it Progress in nonequilibrium QFT I and II},
  hep-ph/0302210 and hep-ph/0410330.
 
\bibitem{Koksma:2010zi}
  J.~F.~Koksma, T.~Prokopec and M.~G.~Schmidt,
  Annals Phys.\  {\bf 325} (2010) 1277.

\bibitem{workinprogress}
F. Gautier, J. Serreau, work in progress.

\bibitem{Blaizot:1999ip}
  J.~P.~Blaizot, E.~Iancu, A.~Rebhan,
  Phys.\ Rev.\ Lett.\  {\bf 83} (1999) 2906.

\bibitem{ZJ}
 J. Zinn-Justin, {\it Quantum Field Theory and Critical Phenomena}, Clarendon Press - Oxford, 4th editon (2002).
 
\bibitem{Berges:2001fi}
  J.~Berges,
  Nucl.\ Phys.\  A {\bf 699} (2002) 847;

\bibitem{Berges:2002wr}
  J.~Berges, S.~Bors\'anyi, J.~Serreau,
  Nucl.\ Phys.\  B {\bf 660} (2003) 51.

\bibitem{Cooper:2002qd}
  F.~Cooper, J.~F.~Dawson and B.~Mihaila,
  Phys.\ Rev.\  D {\bf 67} (2003) 056003.

\bibitem{Juchem:2003bi}
  S.~Juchem, W.~Cassing and C.~Greiner,
  Phys.\ Rev.\  D {\bf 69} (2004) 025006.

\bibitem{Arrizabalaga:2005tf}
  A.~Arrizabalaga, J.~Smit and A.~Tranberg,
  Phys.\ Rev.\  D {\bf 72} (2005) 025014.

\bibitem{Tranberg:2008ae}
  A.~Tranberg,
  JHEP {\bf 0811} (2008) 037.
  
\bibitem{Cooper:1994hr}
  F.~Cooper, S.~Habib, Y.~Kluger, E.~Mottola, J.~P.~Paz and P.~R.~Anderson,
  Phys.\ Rev.\  D {\bf 50} (1994) 2848.

\bibitem{Lebellac} M. Le Bellac {\it Thermal Field Theory}, Cambridge university Press (1996).

\bibitem{Glauber:1963tx}
  R.~J.~Glauber,
  Phys.\ Rev.\  {\bf 131} (1963) 2766.

  

\end{thebibliography}
\end{document}